\documentclass[a4,twocolumn,showpacs,amsmath,pra,amssymb,superscriptaddress,floatfix]{revtex4-1} 
\usepackage[colorlinks=true,urlcolor=blue,citecolor=blue,linkcolor=blue]{hyperref}
\usepackage[T1]{fontenc}
\usepackage[latin9]{inputenc}
\usepackage{times}
\usepackage{color}
\usepackage{xspace}
\usepackage{amssymb,amsmath}
\usepackage{amsbsy}
\usepackage{graphicx}
\usepackage{epstopdf}
\usepackage{float}
\usepackage[T1]{fontenc}
\usepackage{pslatex}
\usepackage{contour}
\usepackage{color}
\usepackage[usenames,dvipsnames]{xcolor}
\newcommand{\ket}[1]{|#1\rangle}
\usepackage[colorlinks=true,urlcolor=blue,citecolor=blue,linkcolor=blue]{hyperref}
\usepackage{color}
\definecolor{blaa}{RGB}{153,153,255}
\definecolor{filtered}{RGB}{153,0,51}

\usepackage{amsmath}
\usepackage{amssymb}
\usepackage{epstopdf}
\usepackage[T1]{fontenc}
\usepackage{times}

\newcommand{\eref}[1]{Eq.~(\ref{#1})}
\newcommand{\sref}[1]{Sec.~\ref{#1}}
\newcommand{\fref}[1]{Fig.~\ref{#1}}

\newcommand{\aref}[1]{App.~\ref{#1}}

\newcommand{\Fref}[1]{Figure~\ref{#1}}

\definecolor{raw}{RGB}{255,177,100}
\definecolor{filtered}{RGB}{153,0,51}

\begin{document}
\title{Dispersive optical detection of magnetic Feshbach resonances in ultracold gases}
\author{Bianca J. Sawyer} \email{bianca.j.sawyer@postgrad.otago.ac.nz}

\author{Milena S.J. Horvath}
\affiliation{Department of Physics, QSO---Centre for Quantum Science, and Dodd-Walls Centre for Photonic \& Quantum Technologies, University of Otago, Dunedin, New Zealand}
\author{Eite Tiesinga}
\affiliation{Joint Quantum Institute and Center for Quantum Information and Computer Science, National Institute of Standards and Technology and University of Maryland, Gaithersburg, Maryland 20899, USA}
\author{Amita B. Deb}
\author{Niels Kj{\ae}rgaard} \email{nk@otago.ac.nz}
\affiliation{Department of Physics, QSO---Centre for Quantum Science, and Dodd-Walls Centre for Photonic \& Quantum Technologies, University of Otago, Dunedin, New Zealand}
\date{\today}

\begin{abstract}

Magnetically tunable Feshbach resonances in ultracold atomic systems are chiefly identified and characterized through time consuming atom loss spectroscopy. We describe an off-resonant dispersive optical probing technique to rapidly locate Feshbach resonances and demonstrate the method by locating four resonances of $^{87}$Rb, between the $|\mathrm{F} = 1, \mathrm{m_F}=1 \rangle$ and $|\mathrm{F} = 2, \mathrm{m_F}=0 \rangle$ states. Despite the loss features being $\lesssim0.1$~G wide, we require only 21 experimental runs to explore a magnetic field range $>18$~G, where $1~\mathrm{G}=10^{-4}$~T. The resonances consist of two known $s$-wave features in the vicinity of 9~G and 18~G and two previously unreported $p$-wave features near 5~G and 10~G. We further utilize the dispersive approach to directly characterize the two-body loss dynamics for each Feshbach resonance.

\end{abstract}

\maketitle

\section{Introduction}
Trapped ultracold gases have long been utilized as a highly controllable experimental testbed for the investigation of exciting and fundamental quantum phenomena such as matter wave interference \cite{Cronin2009}, superfluidity \cite{Desbuquois2012}
and the BEC-BCS crossover \cite{Randeria2014}. The standard technique for observing a cold atomic sample is time-of-flight absorption imaging where the entire sample is released from the trapping potential and illuminated with resonant laser light, projecting a shadow onto a charge coupled device (CCD) camera that gives the two-dimensional density distribution of the sample \cite{Ketterle1999}. Conventional absorption imaging provides valuable information on the spatial distribution and internal quantum state of atoms. However, during the process the atoms are released from the confining potential and undergo a strong resonant interaction with the probe laser light, which heats up and destroys the sample. This allows for acquisition of just one data point per experimental run, and has motivated the development of probing methods using off-resonant light to reduce the spontaneous scattering of photons away from the probe beam, such as dispersive dark-ground imaging \cite{Andrews1996}, phase contrast imaging \cite{Andrews1997,Wigley2016}, and Faraday imaging \cite{Kaminski2012, Gajdacz2013, Gajdacz2016}.

For many applications the object of interest is not the spatial distribution of the cloud, but the temporal evolution of the atomic population within a given probing volume (an integral over the spatial distribution). For this class of measurements, a single photodiode would suffice for efficient data collection. This approach, in particular in conjunction with off-resonant probe light, has for example been used to monitor breathing \cite{Petrov2007} and spatial center-of-mass oscillations \cite{Kohnen2011} of atomic samples, Rabi oscillations between hyperfine states \cite{Chaudhury2006, Bernon2011,Deb2013}, phase-space dynamics of spinor condensates \cite{Liu2009} and Larmor precession \cite{Isayama1999}. While the spatial information imprinted on the probe laser beam is not retained, a photodiode provides an effective means for collecting high-bandwidth real-time temporal information during dynamical processes, for example, recording Rabi oscillations at a sample rate of $\simeq1$~MHz \cite{Windpassinger2008}.

Feshbach resonances fall perfectly into the category of phenomena that can be efficiently explored through integrated dispersive measurements; the atom loss dynamics for a trapped gas driven by a Feshbach resonance is usually characterized via the total atom number, disregarding spatial information. The study of Feshbach resonances has remained an active field for more than two decades \cite{Tiesinga1992, Inouye1998, Maier2015}, in particular because such resonances provide a powerful tool for tuning the scattering properties of atoms through external fields. Recent experiments include controlling the two-body Feshbach losses using electromagnetically induced transparency \cite{Jagannathan2016}, investigation of $p$-wave Feshbach resonances in $^{6}$Li \cite{Nakasuji2013}, and an interorbital Feshbach resonance in $^{173}$Yb \cite{Hoefer2015}. There have also been recent proposals to produce Feshbach resonances using rf fields \cite{Papoular2010,Owens2016} and Rydberg molecular states \cite{Sandor2016}.

In this paper we demonstrate the use of hyperfine state sensitive dispersive probing --- measurement of the quantum state dependent phase shift acquired by an off-resonant probe beam as it passes through a sample --- for efficiently locating and exploring the loss dynamics in connection to Feshbach resonances. In particular we consider $^{87}$Rb, which is one of the most prolifically utilized species in cold atom experiments worldwide, motivating the quest for a thorough understanding of its scattering properties, including details of its landscape of Feshbach resonances \cite{Tojo2010,Regal2003}. We explore the magnetic field dependent collisional loss due to interactions between atoms in the $5^2S_{1/2}|\mathrm{F} = 1, \mathrm{m_F}=1 \rangle \equiv |1,1\rangle$ and $5^2S_{1/2}|\mathrm{F} = 2, \mathrm{m_F}=0 \rangle \equiv |2,0\rangle$ hyperfine states, and identify four resonances in the range $0$~G to $18$~G \cite{noteTvsG,Horvath2017}. These consist of two previously observed $s$-wave features \cite{Kaufman2009}, and two $p$-wave features that have not been reported in prior experiments.

\section{Experimental setup}
\label{Sec:StatePrep}

\begin{figure*}[!htbp]
	\centering
	\includegraphics[trim={0cm 0.3cm 0cm 0.8cm},width=\textwidth]{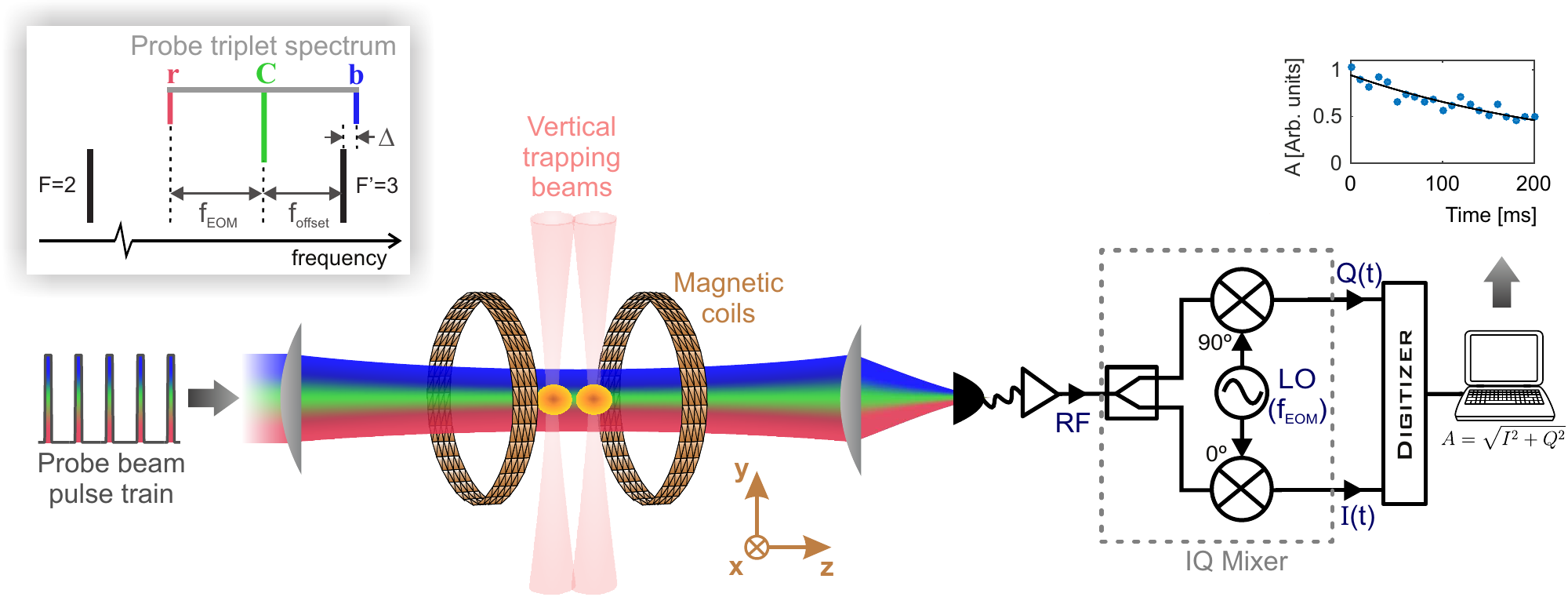}
	\caption{(Color online) Schematic of the experimental setup, showing propagation of the pulsed trichromatic dispersive probing beam through the sample of optically-trapped $^{87}$Rb, followed by the demodulation electronics which extract phase shift information from the photo-detected signal. A typical post-processed data set and fit is shown in the plot to the far right. Note that the horizontal trapping beam, which completes the optical tweezer system, is co-propagating with the dispersive probe but is not shown here. Inset in upper left corner: the trichromatic frequency spectrum of the probe, consisting of a carrier (C), and red (r) and blue (b) sidebands. The probe is shown relative to the $^{87}\mathrm{Rb}$ $(5^2\mathrm{S}_{1/2},~ \mathrm{F}=2)\rightarrow (5^2\mathrm{P}_{3/2},~ \mathrm{F}'=3)$ absorption line (not to scale).}
	\label{fig:Setup}
\end{figure*}

\subsection{Ultracold sample production}
We begin the experiment by producing an ultracold sample of $^{87}$Rb atoms in the $\ket{2,2}$ hyperfine Zeeman substate, using a standard laser-cooling apparatus \cite{Sawyer2012}. We then transfer the sample from a magnetic trap into a double-well potential formed by two crossed-beam far-off-resonant dipole traps \cite{Roberts2014} and evaporatively cool to a temperature of $1.4~\mu$K by lowering the optical power of the horizontal confinement beam, as detailed in \cite{Deb2014}. The sample now consists of two closely-spaced ellipsoidal atom clouds positioned along the $z$-axis. The double-well potential facilitates efficient loading of atoms from the elongated magnetic potential and provides the benefit of increased peak density over an elongated single-well potential, which increases the rate of Feshbach losses. Each well is characterized by axial and radial trapping frequencies $(\omega_x,\omega_y, \omega_z) = 2\pi × (243, 132, 156)$~Hz, where the coordinate system is defined in \fref{fig:Setup}.

\subsection{Applied magnetic field}
The magnetic field at the position of the atomic cloud is controlled by a coil pair arranged in the Helmholtz configuration, and points along the $z$-axis. This field, $\mathbf{B}=B_z\hat{\mathbf{z}}$, defines a quantization axis and lifts the degeneracy of the Zeeman sub-levels. Since the Feshbach resonances in $^{87}$Rb are narrow (all the known resonances are $\lesssim200$~mG wide), we use a current supply with 10~$\mu$A/A stability to drive the Helmholtz coils. An arbitrary waveform generator controls the current supply, and the generated magnetic field has a stability better than $0.2$~mG and has been calibrated using Rabi spectroscopy to an accuracy of $\simeq3$~mG. The experimental setup is summarized schematically in \fref{fig:Setup}.

\subsection{Quantum state preparation}
\label{Sec:QSP}
\begin{figure}[!htbp]
	\centering
	\includegraphics[trim={0cm 0cm 0cm 0cm}, width=\linewidth]{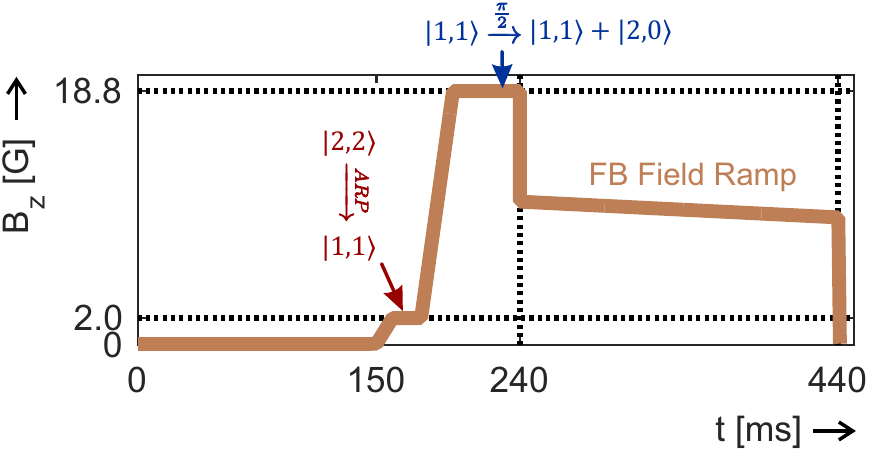}
	\caption{(Color online) Typical magnetic field profile during an experimental run, indicating the timing of the two state preparation stages; adiabatic rapid passage (ARP) and a resonant $\pi/2$-pulse ($\frac{\pi}{2}$).}
	\label{fig:StatePrep}
\end{figure}

We convert our $|2,2\rangle$ ultracold $^{87}$Rb sample into a nearly 50-50 mixture of the $|2,0\rangle$ and $|1,1\rangle$ quantum states using microwave-frequency transitions. First, we use a frequency sweep across the $|2,2\rangle \leftrightarrow |1,1\rangle$ resonance to transfer the population by adiabatic rapid passage (ARP) \cite{Metcalf2016} to $|1,1\rangle$, in the presence of a small homogeneous bias field $B_{z} = 2.0$~G. We ensure purity of our sample by removing any atoms remaining in the $F=2$ multiplet with a 1~ms optical clearing pulse, resonant with the $\{5^2S_{1/2},~F=2\} \rightarrow \{5^2P_{3/2},~F'=3\}$ optical transition. The field is then ramped up to 18.8~G over 10~ms and a $\pi/2$-pulse, resonant with the $|1,1\rangle \rightarrow |2,0\rangle$ transition, is used to prepare the sample in a superposition of the $|2,0\rangle$ and $|1,1\rangle$ states, with $N \simeq 2.3 \times 10^6$ atoms at $1.4~\mu$K. A typical magnetic field profile over the course of our experiment is shown schematically in \fref{fig:StatePrep}, indicating the two state preparation stages and the magnetic field ramp used to scan for Feshbach resonances.

Before beginning our investigation of Feshbach dynamics we hold the sample in the trap for a further 12~ms, which exceeds the coherence time of the system, so the resulting sample can be treated as a 50--50 mixture of atoms in the $|2,0\rangle$ and $|1,1\rangle$ states. We obtain two samples with $1/e$ Gaussian radii of $(\sigma_x, \sigma_y, \sigma_z) = (11, 20, 17)~\mu$m and a peak density of $n_0 \simeq 2\times10^{19}~\mathrm{m}^{-3}$ in each of the two atomic substates.

\subsection{Dispersive probe}

The off-resonant dispersive probe beam is locked to a beat note with a reference laser, to within 4~MHz (the 3~dB width of the beat note). This puts the frequency of the laser $f_\text{offset}=-3.30$~GHz below the $\mathrm{F}=2\rightarrow \mathrm{F}'=3$ transition of the D2 line. By passing the probe beam through a fiber electro-optic phase modulator we produce a trichromatic spectrum with 1$^{\mathrm{st}}$-order sidebands at $\pm f_{\mathrm{EOM}} = \pm3.700$~GHz, where the carrier component (C) has an optical power of $\simeq13~\mu$W and each of the two first-order sidebands contain $\simeq1~\mu$W (higher order sidebands have negligible power). The down-shifted sideband (r) is far-red detuned $\Delta =f_{\mathrm{offset}} - f_{\mathrm{EOM}}=-7.00$~GHz from the $2\rightarrow3'$ transition, while the up-shifted (b) sideband has a comparatively small blue detuning of $\Delta =f_{\mathrm{offset}} + f_{\mathrm{EOM}}=+400$~MHz. This achieves a common-path interferometric probe triplet spectrum \{r,C,b\} as illustrated schematically in the inset of \fref{fig:Setup} with respect to the $2\rightarrow3'$ probing transition. The transitions to $\mathrm{F}'=1$ and $\mathrm{F}'=2$ levels are also allowed, but are much weaker \cite{Steck2001b} and can be ignored.

The values of $f_\text{offset}$ and $f_{\mathrm{EOM}}$ determine the positions of all three dispersive probe components relative to the probing transition, and are chosen with a number of considerations in mind. Notably, we maximize the detuning of the C and r components within the technical limitations imposed by the detector and amplifier bandwidths, to avoid significant interaction with the atomic sample. We also position the C and r components on the opposite side of resonance to the b sideband, in order to reduce the net light shift. We note that the probing scheme is insensitive to atoms in the $F=1$ state, as the $F=1 \rightarrow F'$ transition is far-off-resonant for all probe triplet components.
	
The probe beam is modulated to create a train of probe pulses, which are linearly polarized along the $x$-axis and propagated along the $z$-axis, being focused to a $28~\mu$m waist centered on a sample. As the probe triplet passes though the atomic cloud, the blue frequency component acquires a phase shift dependent upon the $F=2$ population of the cloud. The beam is then focused onto a 4.2~GHz bandwidth fiber-coupled ac photodetector, where the three frequency components combine to produce a heterodyne signal at frequency $f_{\mathrm{EOM}}$ \cite{Bjorklund1980}. Following an amplification stage we pass the signal through a passive I-Q mixer, extracting the in-phase (I) and quadrature (Q) components by demodulating with a frequency $f_{\mathrm{EOM}}$. The output of each mixer port is sampled at a rate of $20~\mu\text{s}^{-1}$ with a 16-bit digitizer. At the instance of each probe pulse, the digitized I and Q components are numerically integrated over and summed in quadrature to give the dispersive signal, $A(t) = \sqrt{I^2(t)+Q^2(t)}$. In the regime where $\Delta=400$~MHz is much larger than the 6~MHz natural linewidth of the $^{87}$Rb D2 line, the dispersive signal is proportional to the phase shift acquired ($A(t)\propto\phi$) and thus to the $|2,0\rangle$ population \cite{Lye1999}, provided the geometry of the sample does not change (see \aref{Sec:heatingDP}). An example processed data set is shown in the upper right corner of \fref{fig:Setup}.

\section{Locating Feshbach resonances}

Following preparation of the sample we sweep the magnetic field down linearly from a series of starting magnetic field values at a constant rate of $-5.83$~mG/ms, with each sweep covering a range of 1.17~G over 200~ms (\fref{fig:StatePrep} shows a typical magnetic field sweep.) During each sweep we monitor the dynamics of the atomic population in the $|2,0\rangle$ state dispersively, with a train of 21 light pulses at intervals of 10~ms. Each pulse has a duration of 600~ns and contains $\simeq 3\times 10^6$ photons in the probing (b) sideband.

\begin{figure}[!htbp]
	\centering
	\includegraphics[trim={0cm 0cm 0cm 0cm}, width=\columnwidth]{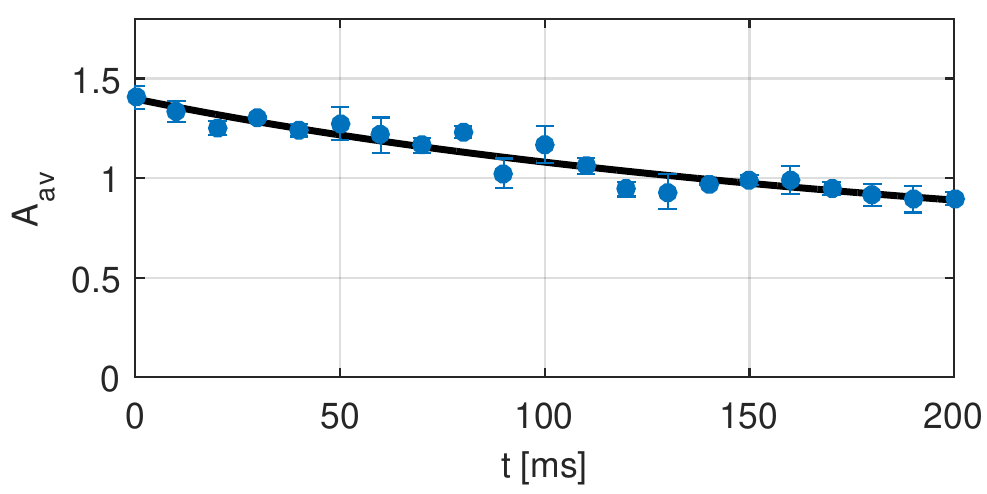}
	\caption{(Color online) Reference dispersive signal, acquired at constant off-resonance magnetic field (blue circles), and a fit to the model presented in \aref{Sec:modelDecay} (solid black line). Error bars represent uncertainties of one standard deviation.}
	\label{fig:RefShot}
\end{figure}

To determine whether a decrease in the dispersive signal acquired during a magnetic field sweep is indicative of a Feshbach resonance, we also measure the background atom losses in a constant off-resonance magnetic field $B_z=18.8$~G. This reference signal, averaged over three runs, is shown in \fref{fig:RefShot} (gray circles) alongside a fit to the model in \aref{Sec:modelDecay}. 

\begin{figure*}[!htbp]
	\centering
	\includegraphics[trim={0cm 0cm 0cm 0cm}, width=\textwidth]{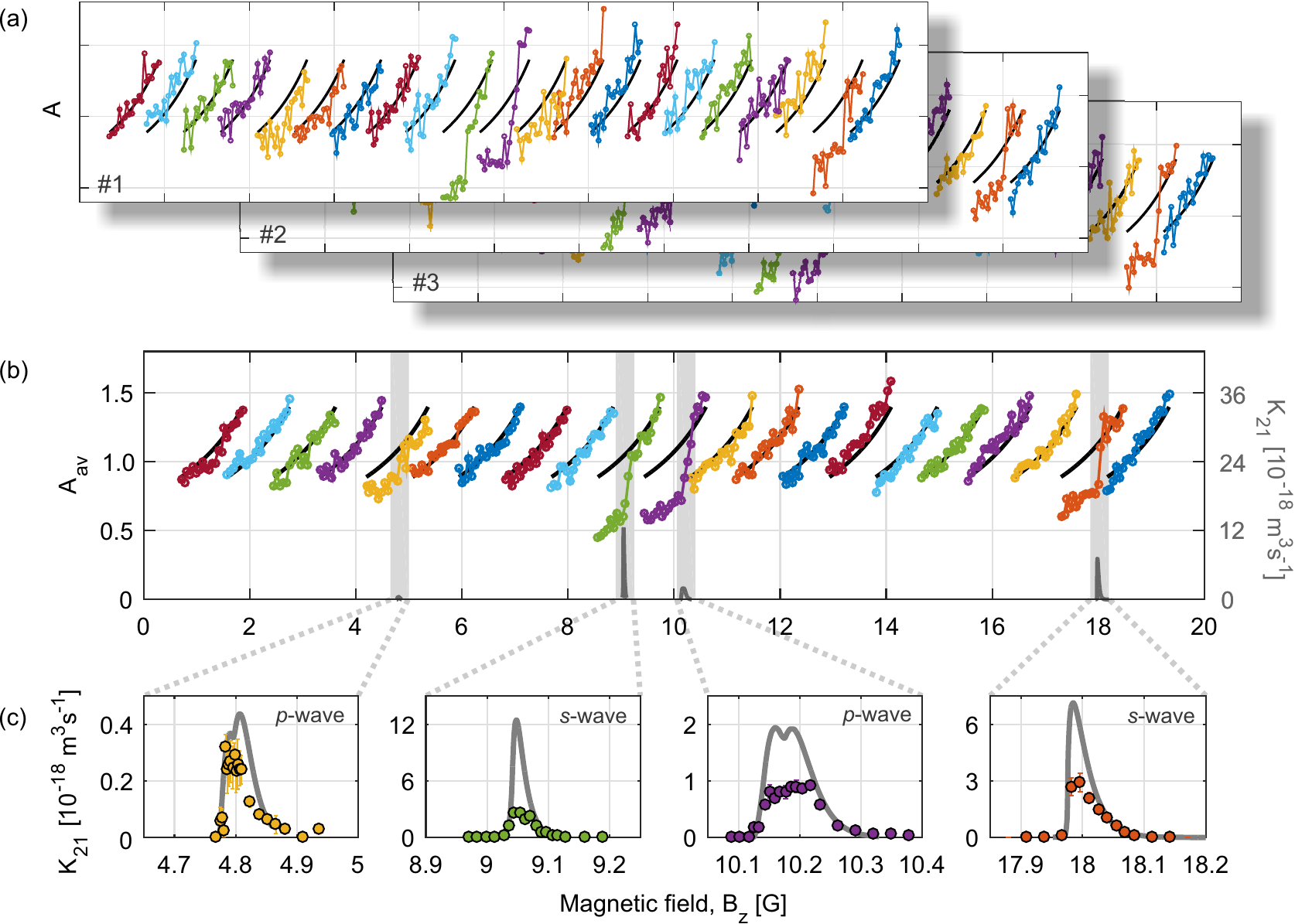}
	\caption{(Color online) (a) Three data sets showing the single-experiment dispersive signal ($A$, arbitrary units) for 21 consecutive and overlapping magnetic field sweeps, covering the range $19.4~\mathrm{G}-0.7$~G. The black line is a reference for comparison, as explained in the main text. (b) Left-hand axis: Dispersive signal averaged over the three experimental runs for each sweep, $A_{\mathrm{av}}$. Right-hand axis: Theoretically predicted two-body loss rate coefficients ($K_{21}$, gray lines below dispersive data), calculated for a thermal ensemble of atoms at 1.4~$\mu$K, about each of the four identified Feshbach resonances. (c) Zoomed-in view of the theoretically predicted $K_{21}$ coefficients (gray lines), alongside the corresponding experimentally measured $K_{21}$ coefficients (colored markers). One standard deviation uncertainties are indicated by error bars, which in most cases do not extend beyond the plotted point size.}
	\label{fig:MainData}
\end{figure*}

\Fref{fig:MainData}(a) presents the results of 21 magnetic field sweeps, which collectively cover the range 19.4~G to 0.7~G and overlap by 25~\% at each edge. The dispersive data set obtained for each sweep is plotted in sequence (colored circles) and superimposed on a fit to the reference signal (black line), with the time axis recalibrated to match each of the 21 magnetic field sweeps, for easy comparison of the two signals. For several sweeps the loss signal deviates from the background trace, indicating possible Feshbach resonances. This is reinforced by \fref{fig:MainData}(b), which shows the average of three repeated measurements for each magnetic field range. There are four clear steps in the signal, shaded gray to indicate Feshbach resonances at approximately 5~G, 9~G, 10~G and 18~G. By repeating the experiment about these four values with a pure sample of each component state ($|2,0\rangle$ and $|1,1\rangle$) separately, we verified that the Feshbach resonances observed all correspond to the mixed-spin entrance channel.

	\begin{figure}[!htbp]
	\centering
	\includegraphics[trim={0cm 0cm 0cm 0cm}, width=\linewidth]{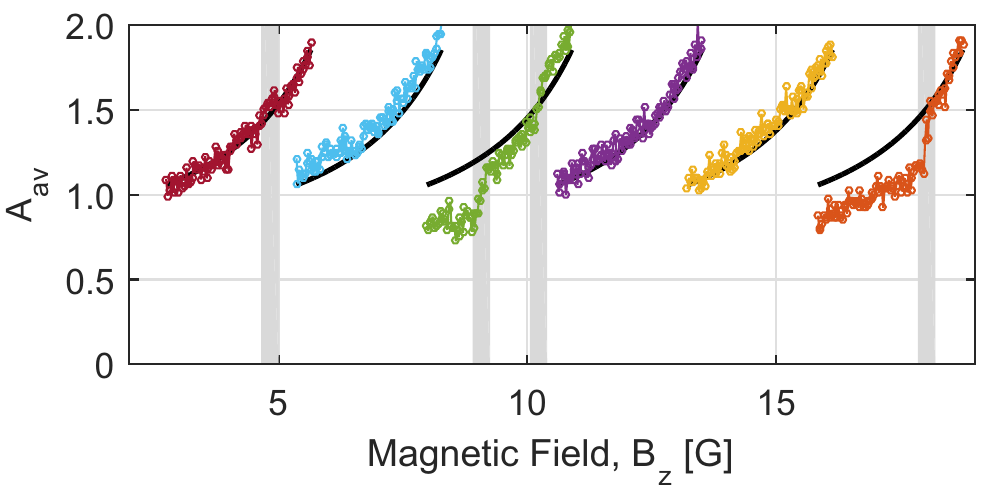}
	\caption{(Color online) Dispersive signal ($A_{av}$, averaged over three experimental runs) recorded during six overlapping magnetic field sweeps. The black lines indicate a reference data set, and the shaded gray boxes indicate regions in which we know there is a Feshbach resonance.}
	\label{fig:6panels}
	\end{figure}

Background losses in the sample limit the total duration of each magnetic field sweep or hold time. The atomic density of the cloud reduces by $40~\%$ over a 200~ms period even in the absence of Feshbach interactions, making density-dependent Feshbach loss more difficult to detect. We require 21 consecutive sweeps to cover the full $>18$~G range at a rate low enough to dispersively detect losses due to the weak resonance near $5$~G. If we are only interested in stronger features we can increase the sweep rate significantly, requiring fewer experimental runs to cover the same range. An example data set is shown in \fref{fig:6panels} where we used just 6 sweeps, each 2.9~G wide, to cover a 16~G range. Shaded gray regions indicate where we expect to see Feshbach resonances, based on our investigation above. Steps in the signal are evident at the $\simeq18~\mathrm{G}, \simeq10$~G and $\simeq9$~G Feshbach resonances, but there is no clear evidence of loss near $5$~G.


\section{Measurement of decay coefficients}

\begin{figure}[!htbp]
	\centering
	\includegraphics[trim={0cm 0.0cm 0cm 0.0cm}, width=\linewidth]{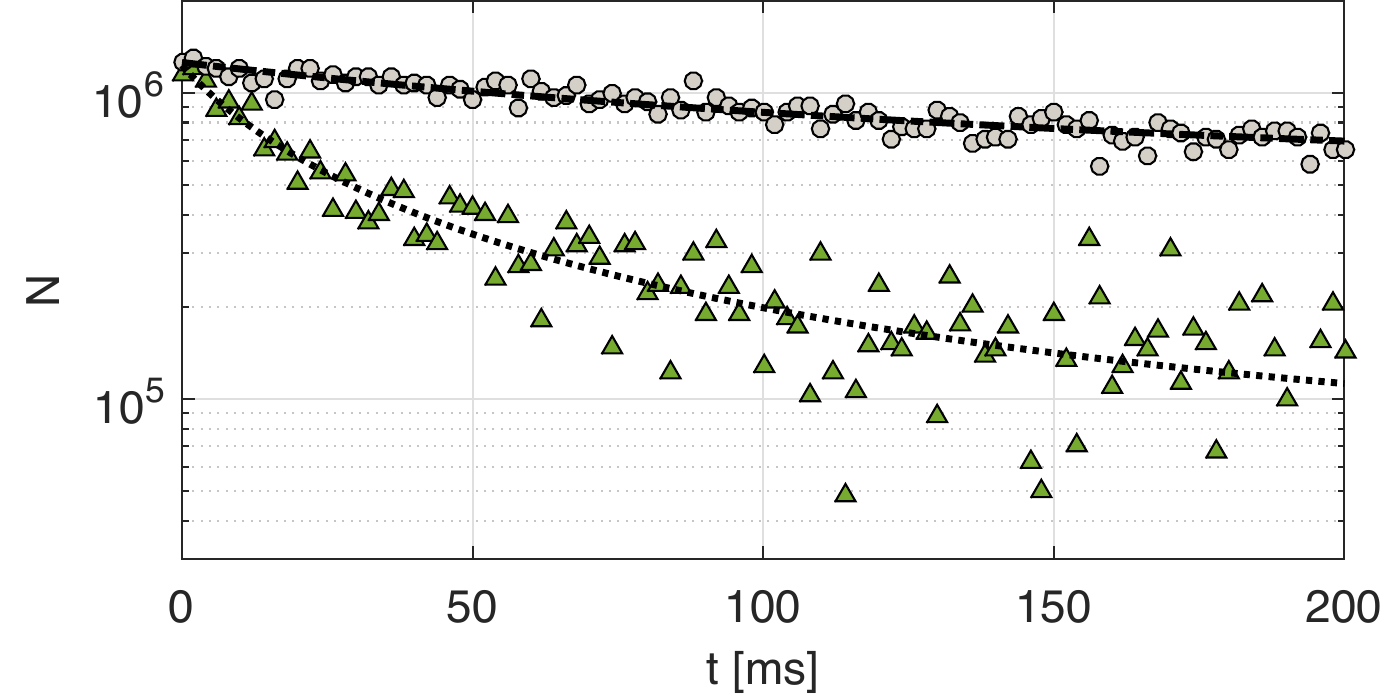}
	\caption{(Color online) Atomic loss data acquired via dispersive probing as we hold the sample for 200~ms off-resonance (9.158~G, gray circles) and near-resonance (9.062~G, green triangles). Fits to the model described in \aref{Sec:modelDecay} are shown by a dash-dotted and dotted line, respectively.}
	\label{fig:Decay}
\end{figure}

We further characterize the four observed Feshbach resonances by dispersively measuring the 2-body loss rate coefficient, $K_{21}$, which describes the rate of enhanced losses near a Feshbach resonance, as a function of magnetic field. The state preparation sequence for the measurement of $K_{21}$ loss rate coefficients is almost identical to that in \sref{Sec:QSP} but with one subtle difference. Because we need high magnetic field stability to precisely characterize narrow loss features, we carry out the $\pi/2$-pulse at a magnetic field $<2$~G above the resonance of interest (at 18.8~G, 10.9~G, 10.9~G and 5.4~G for each resonance respectively). This minimizes ringing of the field when we decrease it to a fixed field value near resonance during the $K_{21}$ investigations, while avoiding atom loss due to Feshbach dynamics during the state preparation. Now rather than sweeping the magnetic field downward, we hold it at a constant value for 200~ms while probing the $|2,0 \rangle$ component of the cloud dispersively with 600~ns pulses at intervals of 2~ms \footnote{For the data about the 18~G resonance, the probing interval is 5~ms. This does not result in any differences in the analysis of the data, but gives data of lower resolution.}, hence following the evolution of atomic population in real time. The dispersive signal is converted to absolute atomic population $N$ using a calibration based on absorption images (see \aref{Sec:modelDecay}). In \fref{fig:Decay} we show example data sets for the atomic population decay at an on-resonance and an off-resonance magnetic field about the Feshbach resonance near 9~G. The off-resonant case displays an exponential decay, while the loss in the on-resonant case is much faster and non-exponential. \Fref{fig:Decay} also presents the result of fitting a nonlinear model to the data that captures both one- and two-body background losses and two-body Feshbach loss features (see \aref{Sec:modelDecay}). The dispersive decay measurements are carried out in the vicinity of each resonance and values of the $K_{21}$ coefficient are extracted from the fits to data, averaged over three data sets for each field value, and plotted in \fref{fig:MainData}(c).

The theoretically predicted $K_{21}$ coefficients for a thermal ensemble of atoms at $1.4~\mu$K are indicated by gray lines in \fref{fig:MainData}(c). These are obtained from numerical coupled-channels calculations based on a Hamiltonian of a homonuclear pair of ground state $^2$S $^{87}$Rb atoms with nuclear spin $i=3/2$ \cite{Stoof1988,Leo2000,Strauss2010} that includes atomic hyperfine and Zeeman interactions, the isotropic X$^1\Sigma_g^+$ and a$^3\Sigma_u^+$ Born-Oppenheimer potentials, the centrifugal potential with partial wave $\vec \ell$, as well as the anisotropic electronic magnetic dipole-dipole and second-order spin-orbit interactions. Spectroscopically-accurate Born-Oppenheimer potentials and the parametrization of the second-order spin-orbit interaction are taken from Ref.~\cite{Strauss2010}. The Hamiltonian conserves the sum of all spin and angular momentum projection quantum numbers and parity. Hence, even and odd partial waves remain uncoupled. As the two anisotropic interactions for $^{87}$Rb are weak it suffices to include all $\ell=0$ and 2 and $\ell=1$ and 3 channels for our $s$-wave and $p$-wave Feshbach resonances, respectively. Elastic and inelastic rate coefficients near the Feshbach resonances are first computed as a function of collision energy and then thermally averaged using energies up to ten times the temperature. Taking the resonance positions to be at the local maxima in the calculated $K_{21}$ values, we get predicted $s$-wave peaks at $9.048$~G and $17.985$~G and p-wave doublets at \{$4.792$~G, $4.806$~G\} and \{$10.160$~G, $10.195$~G\}.

Agreement of the four experimentally determined resonance positions with theory is good and the expected qualitative characteristics are present in the data. There is a clear asymmetry in the $K_{21}$ coefficient, with the tail trailing out toward higher values of magnetic field due to thermal broadening and there is also some evidence of the doublet structure of the $p$-wave Feshbach resonance near 10~G, which manifests due to the dipole-dipole interaction having different values depending on partial-wave projection, $|m_l| = \{0,1\}$ \cite{Ticknor2004}. The $K_{21}$ values shown in \fref{fig:MainData}(c) are based on the assumption that the temperature is fixed at $1.4~\mu$K throughout the 200~ms hold time, and for the resonance near $4.8$~G, where the loss rate coefficient is small, we get good quantitative agreement with theory. For the other three resonances we get agreement in the wings, but the inferred $K_{21}$ values are significantly lower than the theoretical predictions at the peak of each feature. We attribute this discrepancy to ``anti-evaporative'' heating of the sample \cite{Chin2016}, as Feshbach loss occurs preferentially from high density (low energy) regions of the sample, and the kinetic energy of the ejected atoms can be partially transferred to the thermal energy of the sample via collision with other particles during their escape \cite{Regal2007}.

To gain further insight into the effect of heating, we acquire an absorption image following the 200~ms hold time for each data set. From this it is apparent that larger Feshbach losses lead to a higher temperature increase, consistent with our finding of a larger discrepancy at the peak of the loss features. Indeed, we find the predicted values of $K_{21}$ to lie closer to a modeled $K_{21}$ based on the final temperatures (for details see \aref{Sec:heating}). While heating limits the applicability of our model to accurate measurement of large $K_{21}$ values, it does not reduce the efficacy of the method for the purposes of detecting the locations of Feshbach resonances. A refined model, taking into account the time dependence of the temperature, would be required to properly estimate $K_{21}$ at the peak. This is outside the scope of this work.

The decay measurements fully exploit the potential of the dispersive probe system to considerably speed up data acquisition. To acquire equivalent information for Feshbach loss dynamics using standard time-of-flight absorption imaging, one would require a full experimental sequence ($\approx100$~s in our setup) per data point. In addition, using dispersive probing the dynamics can be monitored on a microsecond timescale with a high bandwidth (up to 1.6~MHz). Such rapid data collection also minimizes effects of drifting background fields and other experiment conditions, reducing sources of systematic error.
	
\section{Conclusion}
In conclusion, we have demonstrated the use of an off-resonant heterodyne optical dispersive probing system to efficiently detect and characterize Feshbach resonances in ultracold $^{87}$Rb. Our measurements of the two previously unreported $p$-wave resonances fill a gap in the rich body of data on the widely used $^{87}$Rb species. The method provides a powerful new tool for mapping out the Feshbach resonances of any pair of substates, and could straightforwardly be extended to any species with a change in optical frequencies, including those with optical Feshbach resonances \cite{Blatt2011}.

Dispersive probing could be further utilized to investigate other types of loss dynamics near a Feshbach resonance, such as three-body losses and the associated Efimov signatures \cite{Ferlaino2011}. This will extend the applicability of the technique to broad resonances, where 2-body inelastic losses may be negligible over $\approx1$~s timescales. Finally, our method may provide an effective tool for the study of coherent atom-molecule oscillations in a BEC \cite{Donley2002}.

\section*{Acknowledgment}
We thank Ryan Thomas for useful discussions.


\appendix
\section{Model for decay data and dispersive signal}
\label{Sec:modelDecay}

In this section we derive a model for the time evolution of a trapped sample population held in a constant magnetic field. We then relate this to the measured dispersive signal, from which we extract the parameters describing the loss dynamics near a Feshbach resonance.

The crossed-beam dipole-trapped $|1,1\rangle$ and $|2,0\rangle$ state populations, $N_1(t)$ and $N_2(t)$, respectively, are modeled by starting from the coupled rate equations
\begin{equation}
\frac{dn_1(\mathbf{r},t)}{dt} = - \Gamma_1 n_1(\mathbf{r},t) - 2K_{11}n_1(\mathbf{r},t)^2 - K_{21}n_1(\mathbf{r},t)n_2(\mathbf{r},t),
\label{Eq:Rate11n}
\end{equation}
and
\begin{equation}
\frac{dn_2(\mathbf{r},t)}{dt} = - \Gamma_2 n_2(\mathbf{r},t) - 2K_{22}n_2(\mathbf{r},t)^2 - K_{21}n_2(\mathbf{r},t)n_1(\mathbf{r},t),
\label{Eq:Rate20n}
\end{equation}
where $n_1(\mathbf{r},t)$ and $n_2(\mathbf{r},t)$ are the atomic densities of the populations at time $t$ and position $\mathbf{r}$ in the crossed-beam dipole trap, $\Gamma_1$ and $\Gamma_2$ are the one-body loss rates due to collisions with molecules or atoms in the background vacuum, and $K_{11}$, $K_{22}$ and $K_{21}$ are the thermally-averaged two-body loss rate coefficients for $\{1,1+1,1\}$, $\{2,0+2,0\}$ and $\{2,0+1,1\}$ interactions, respectively \cite{Chin2010,Beaufils2009} (we exclude three-body recombination processes, which we estimate to be negligible in our system). The factor of two preceding the $K_{11}$ and $K_{22}$ terms in \eref{Eq:Rate11n} and \eref{Eq:Rate20n} arises because each collision event leads to the loss of two $|1,1\rangle$ or two $|2,0\rangle$ atoms, respectively, rather than one of each as in the $K_{21}$ process.

For a 3D Gaussian density profile with width $(\sigma_x,\sigma_y,\sigma_z)$, \eref{Eq:Rate11n} and \eref{Eq:Rate20n} can be integrated over space to give the rate equations for the crossed-beam dipole-trapped sample populations,
\begin{subequations}
	\label{eq:rateeqns1}
	\begin{align}
		\frac{dN_1(t)}{dt} =& - \Gamma_1 N_1(t) - \frac{2K_{11}}{{(2\pi)^{3/2}\sigma_x\sigma_y \sigma_z}} N_1(t)^2\nonumber \\
		&{-} \frac{K_{21}}{{(2\pi)^{3/2}\sigma_x\sigma_y \sigma_z}} N_1(t)N_2(t),
		\label{Eq:Rate11}
	\end{align}
	\begin{align}
		\frac{dN_2(t)}{dt} =& - \Gamma_2 N_2(t) - \frac{2K_{22}}{(2\pi)^{3/2}\sigma_x\sigma_y \sigma_z} N_2(t)^2\nonumber \\
		&{-} \frac{K_{21}}{{(2\pi)^{3/2}\sigma_x\sigma_y \sigma_z}} N_2(t)N_1(t).
	\label{Eq:Rate20}
	\end{align}
\end{subequations}

The cloud widths are temperature dependent, with the relationship given by
\begin{equation}
\sigma_i = \sqrt{\frac{2~k_{\mathrm{B}} T}{m}}\frac{1}{\omega_i},
\label{Eq:sigT}
\end{equation}
where $i = x,y,z$, $k_{\mathrm{B}}$ is the Boltzmann constant, $T$ the temperature of the atoms in the crossed-beam dipole trap, $m$ the mass of a $^{87}$Rb atom and $\omega_i$ the trapping frequency of the dipole trap along direction $i$.

\begin{figure}[b!]
		\centering
		\includegraphics[trim={0cm 0cm 0cm 0cm}, width=\columnwidth]{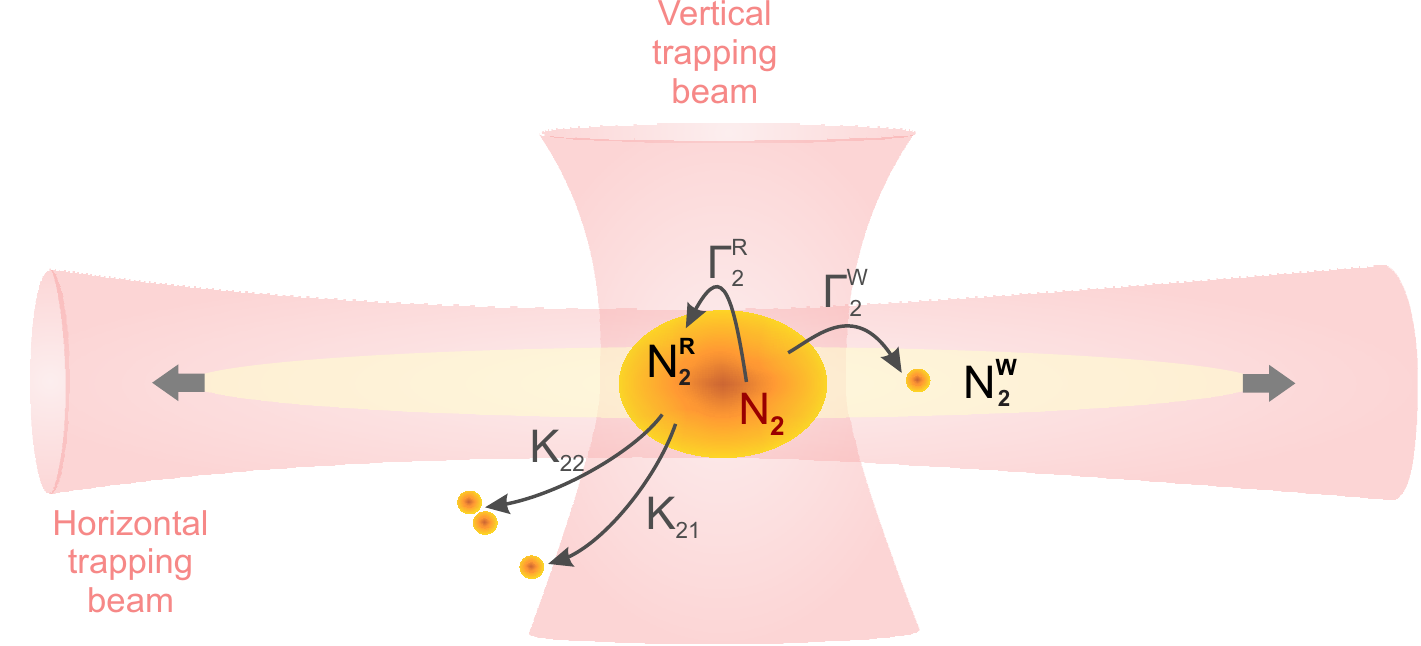}
		\caption{(Color online) Schematic of the three $F=2$ populations that contribute to the dispersive signal. $N_2$ ($\ket{2,0}$ atoms) and $N_2^{\mathrm{R}}$ (a mixture of $\ket{2,1}$ and $\ket{2,-1}$ atoms) are confined by the crossed-beam dipole trap, while $N_2^{\mathrm{w}}$ atoms are confined by the horizontally-propagating trapping potential only, and arrows at each end indicate continued expansion along this waveguide beam. $\Gamma_2^{\mathrm{R}}$ is the rate of optical pumping via Raman processes, $\Gamma_2^{\mathrm{w}}$ is the rate of atom removal from population $N_2$ to population $N_2^{\mathrm{w}}$, and $K_{22}$ and $K_{21}$ are the two-body losses from population $N_2$ to the background vacuum. Note that the density of the waveguide population is much lower than that of the sample, so two-body losses are negligible.}
		\label{fig:DecayPops}
	\end{figure}

The dispersive probe beam causes a small perturbation to the $|2,0\rangle$ component of our sample; a series of 101 dispersive pulses typically results in a $10~\%$ temperature increase and a $20~\%$ population decrease for the experiments in this work, as measured using absorption imaging. It appears that the affected atoms are pushed out of the crossed-trap potential but remain trapped by the horizontal waveguide beam, and a second distinct population of $|2,0\rangle$ atoms, $N_2^{\mathrm{w}}$, accumulates. Because the horizontal waveguide beam propagates coaxially with the dispersive probe beam, this population is still in the path of the dispersive beam. The presence of atoms trapped in the waveguide beam has been confirmed in absorption images at short time-of-flight following dispersive probing of the sample. Additionally, a small number of atoms are expected to undergo Raman transitions via $F'=3$ to the $\ket{2,-1}$ and $\ket{2,1}$ states during the probing sequence (we calculate $\approx13~\%$ in total over a sequence of 101 pulses). This optically pumped population of atoms ($N_2^{\mathrm{R}}$) has the same geometry as the main $\ket{2,0}$ sample and will contribute with near equal weight to the dispersive signal \cite{Steck2001b}. A schematic of the three trapped populations and their associated loss rate coefficients is shown in \fref{fig:DecayPops}.
	
Our dispersive probe measurement scheme is sensitive to all atoms in the $F=2$ ground-state hyperfine manifold, and produces a signal proportional to the $F=2$ state population
\begin{equation}
A = C_2(N_2 + N_2^{\mathrm{R}}) + C_2^{\mathrm{w}} N_2^{\mathrm{w}}+ A_0,
\label{Eq:DPs2}
\end{equation}
where $A_0$ is an offset and $C_2$ and $C_2^{\mathrm{w}}$ are calibration constants. We determine $A_0$ and $C_2$ by comparing dispersive signal amplitudes to the corresponding atom numbers measured using absorption imaging at a temperature of $T=1.4~\mu$K. We can't determine $C_2^{\mathrm{w}}$ because the $N_2^{\mathrm{w}}$ population is small and difficult to measure in absorption images, so we assume that $C_2^{\mathrm{w}}=C_2$. Under this assumption, optical pumping to $\ket{2,-1}$ and $\ket{2,1}$ and atom loss into the waveguide beam both affect the loss dynamics in the same way, reducing the atomic density of $\ket{2,0}$ atoms involved in Feshbach interactions while still contributing to the dispersive signal. We simplify our model by treating the two effects collectively and defining a phenomenological loss rate $\widetilde{\Gamma}_2 = \Gamma_2^{\mathrm{w}} + \Gamma_2^{\mathrm{R}}$, where $\Gamma^w_2$ is the rate at which $|2, 0\rangle$ atoms move from the crossed-dipole trap into the horizontal waveguide due to interaction with the dispersive probe beam, and $\Gamma_2^{\mathrm{R}}$ is the rate of optical pumping to other $m_F$ states.

To describe the effect of the dispersive probe beam on our system we now require a third rate equation,
\begin{equation}
\frac{d\widetilde{N}_2}{dt} = + \widetilde{\Gamma}_2~N_2, 
\label{Eq:RateW}
\end{equation}
and we replace $\Gamma_2$ with $\Gamma_2+\widetilde{\Gamma}_2$ in \eref{Eq:Rate20}. Two-body losses from the $N^{\mathrm{w}}_2$ and $N^{\mathrm{R}}_2$ populations are negligible over the timescale of our experiment, as the atomic densities are very low. From a series of dispersive probe measurements with varying pulse number we have measured a loss rate per probe pulse of 0.011~s$^{-1}$, so for a sequence of 101 pulses, $\widetilde{\Gamma}_2 = 1.1$~s$^{-1}$. 

The dispersive probe does not lead to additional $|1, 1\rangle$ atom losses, as the $|1, 1\rangle$ state is unaffected by the probe. 

\subsection{Fitting background data}

We consider our reference data sets, where the magnetic field is held constant at an off-resonant value $\sim1~\mathrm{G}$ above the Feshbach resonance, and use \eref{Eq:DPs2} to convert the dispersive signal into $F=2$ atomic population. We then fit a model for the temporal evolution of atom number in the absence of two-body Feshbach losses, which is given by the sum of the populations
\begin{equation}
N_{\mathrm{tot}}(t) = N_2(t) + \widetilde{N}_2(t),
\label{Eq:Ntot}
\end{equation}
where the two respective populations are given by the solution to the system of three coupled differential equations
\begin{subequations}
\label{eq:scat_xsec2}
\begin{align}
    \frac{dN_2(t)}{dt}=& - \widetilde{\Gamma}_2 N_2(t) - \frac{2K_{22}}{(2\pi)^{3/2}\sigma_x\sigma_y \sigma_z} N_2(t)^2\nonumber \\
&{-} \frac{K_{21}}{(2\pi)^{3/2}\sigma_x\sigma_y \sigma_z} N_2(t)N_1(t),
\label{Eq:Rate20new}
\end{align}
\begin{align}
   \frac{dN_1(t)}{dt}=& -\frac{K_{21}}{(2\pi)^{3/2}\sigma_x\sigma_y \sigma_z} N_1(t)N_2(t),
\label{Eq:Rate11new} 
\end{align}
\begin{align}
\frac{d\widetilde{N}_2}{dt} = + \widetilde{\Gamma}_2~N_2, 
\label{Eq:RateWnew}
\end{align}
\end{subequations}
with initial conditions $N_2(t=0) = N_1(t=0) = N_0$ and $N_2^{\mathrm{w}}(t=0) = 0$. In deriving this system of equations, we made use of the fact that $(\Gamma_1, \Gamma_2,~ K_{11}n_0) \ll (K_{21}n_0,~ K_{22}n_0)$ for typical values of $n_0$ (the initial single-component peak atomic density) to neglect the terms involving $\Gamma_1, \Gamma_2$ and $K_{11}$ in \eref{Eq:Rate11}, \eref{Eq:Rate20} and \eref{Eq:RateW}. To fit to the off-resonant data sets we set $K_{21}=0$ and used $N_0$ and $K_{22}$ as our fitting parameters. For each of the four data series, $K_{22} \simeq 1.0\times10^{-19}~\mathrm{m}^3\mathrm{s}^{-1}$ (or equivalently, $K_{22}n_0 \simeq 2.0~{\mathrm{s}}^{-1}$).

\subsection{Fitting Feshbach loss data}

Using the values of $K_{22}$ obtained from our off-resonant data sets, we then fitted our near-resonance data sets with \eref{Eq:Ntot} and fitting parameters $N_0$ and $K_{21}$. The extracted values of the two-body loss rate coefficients $K_{21}$ for each magnetic field value are shown in \fref{fig:MainData}(c) of the main text. The maximum two-body peak loss rate of $K_{21}n_0 \simeq 60~\mathrm{s}^{-1}$ was measured for the $\simeq18$~G resonance. Fitting to a simpler model where optical pumping and the losses into the waveguide are neglected (i.e. setting $\widetilde{\Gamma}_2=0$) gives a very similar result, with the extracted $K_{21}$ loss rate coefficients $\leq10~\%$ smaller in magnitude.

\section{Investigation of heating during loss measurements}
\label{Sec:heating}
\begin{figure*}[!htbp]
	\centering
	\includegraphics[trim={0cm 0.5cm 0cm 0cm}, width=\textwidth]{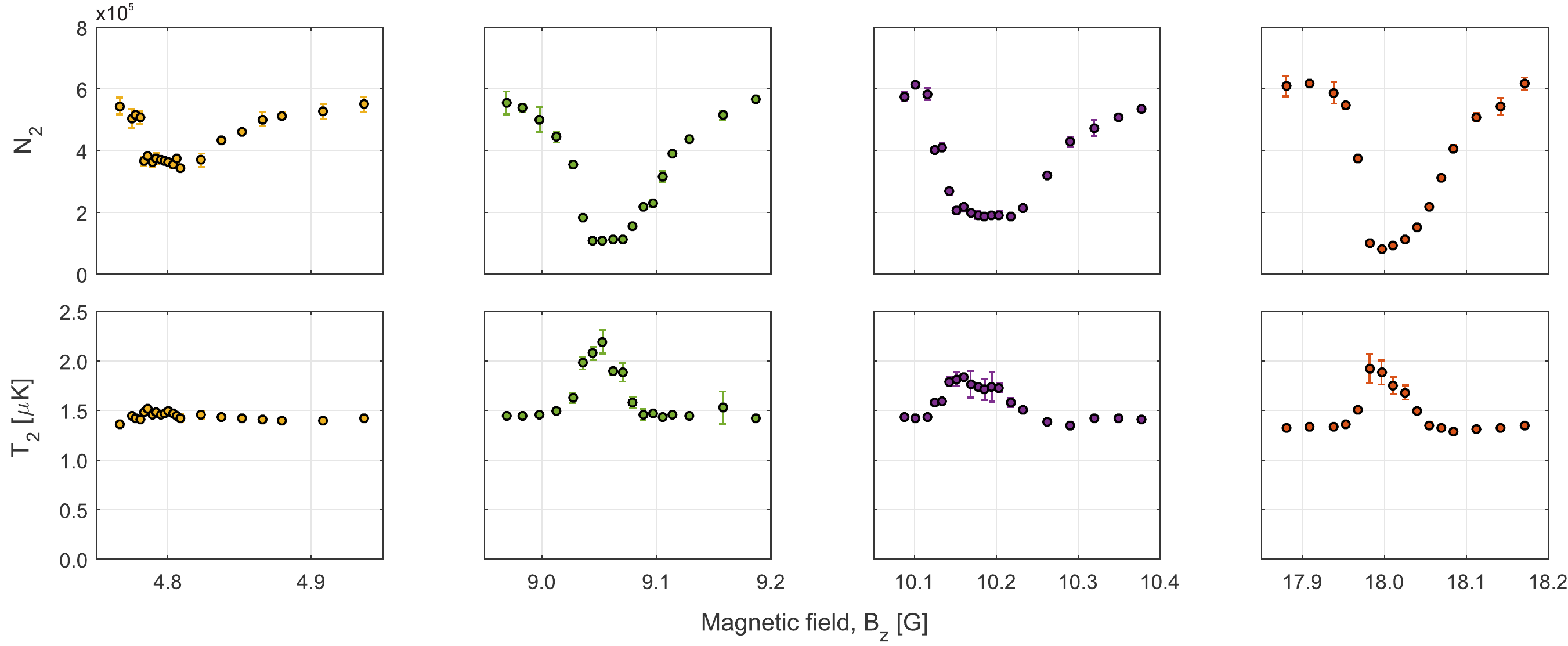}
	\caption{(Color online) Dipole trapped $|2,0\rangle$ population (top row) and temperature (bottom row) derived from time-of-flight absorption images, following a 200~ms hold time at a range of magnetic fields across four Feshbach resonances. One standard deviation uncertainties are indicated by error bars, which in most cases do not extend beyond the plotted point size.}
	\label{fig:CompAIData}
\end{figure*}

We acquired a time-of-flight absorption image immediately following every dispersive measurement. From the images we extracted the spatial distribution of the crossed-beam dipole-trapped $|2,0\rangle$ component following 200~ms of dispersive probing at fixed magnetic field, and calculated the final population and temperature. The measured final $|2,0\rangle$ population in the crossed-trap versus magnetic field is presented in the top row of \fref{fig:CompAIData} and amounts to a set of traditional loss spectroscopy measurements. The resonance positions and qualitative shape of the loss features match with those in \fref{fig:MainData}(c) of the main text, which verifies the validity of our method for analyzing dispersively measured loss dynamics. The measured final temperature versus magnetic field is presented in the bottom row of \fref{fig:CompAIData}, and shows that the temperature of the cloud increases as a result of loss dynamics in the vicinity of a Feshbach resonance. The resonance positions and qualitative shape of the loss features are also mimicked in the temperature data, and we even see the expected doublet-structure near 10~G.

\subsection{Relationship between the decay model and temperature}

\begin{figure*}[!htbp]
	\centering
	\includegraphics[trim={2.6cm 0.3cm 2.7cm 0cm}, width=\textwidth]{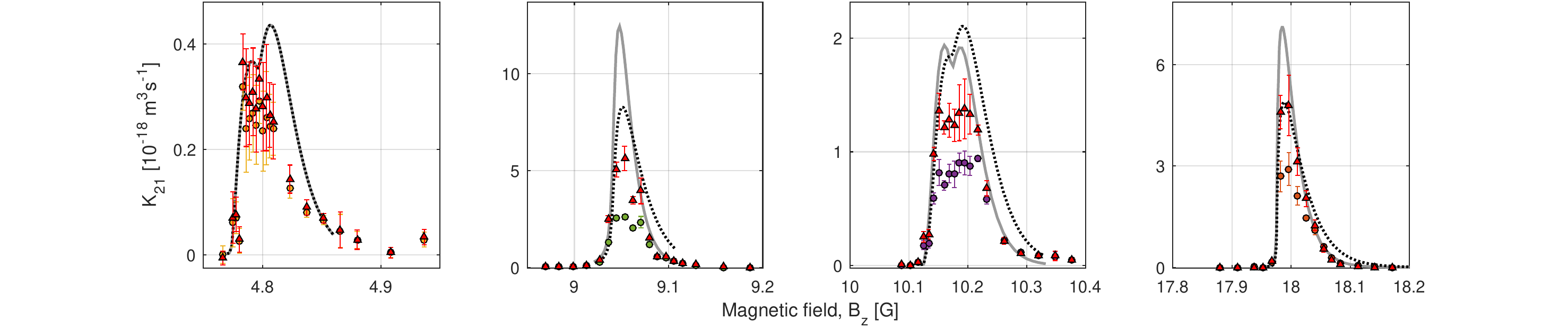}
	\caption{(Color online) Two-body loss rate coefficient, $K_{21}$, for the four identified Feshbach resonances. Colored circles indicate the measured values as in \fref{fig:MainData}(c), and red triangles indicate the measured values based on the expected cloud size at the measured final ensemble temperatures for each data point. In each panel the solid gray line shows the theoretical curve for an ensemble at $1.4~\mu$K while the dotted black line is the theoretical curve for an ensemble at temperatures of $1.6~\mu$K, $2.3~\mu$K, $1.9~\mu$K and $2.0~\mu$K for the resonances in panels a--d respectively. Note that the two curves overlap in the $\simeq5$~G case.}
	\label{fig:Tinv}
\end{figure*}

In deriving the rate equations for the atomic populations, \eref{Eq:Rate11new}, \eref{Eq:Rate20new} and \eref{Eq:RateWnew}, we assumed that the sample had a thermal Gaussian distribution of fixed temperature \cite{Beaufils2009}, and hence that the sample size was constant throughout the loss process. Because the sample heats up during Feshbach interactions it expands according to \eref{Eq:sigT}, and the model presented in \aref{Sec:modelDecay} underestimates the $K_{21}$ loss rate coefficients near resonance. Ideally we need an extended model with time-dependent cloud size $\{\sigma_x(T(t)), \sigma_y(T(t)), \sigma_z(T(t))\}$ to account for the effect of the heating. This would require an understanding of how temperature changes with time, which is not straightforward because heating is correlated with the $K_{21}$ coefficient. While this extension falls outside the scope of our current work, \fref{fig:Tinv} shows the $K_{21}$ values recalculated using the final temperature for each data set in the model detailed in \aref{Sec:modelDecay} (red triangles), rather than the initial temperature of $1.4~\mu$K (also shown for reference, with colored circles). A solid gray line shows the theoretical $K_{21}$ values for an ensemble at $1.4~\mu$K, while the dotted black line is the theoretical curve for an ensemble at temperatures of $1.6~\mu$K, $2.3~\mu$K, $1.9~\mu$K and $2.0~\mu$K, which are the maximum temperatures observed following Feshbach loss about the resonances near $5$~G, $9$~G, $10$~G and $18$~G, respectively.

For data points where the temperature increase during the 200~ms hold time was small, we expect the dispersively measured $K_{21}$ values to match closely with the $1.4~\mu$K (gray solid) theory curve. This is the case in the wings of the 9~G, 10~G and 18~G features, and at most magnetic fields about the 5~G feature, as can be seen in \fref{fig:Tinv}. On the other hand, we expect the dispersively measured $K_{21}$ values to match more closely with the variable upper temperature limit (black dotted) theory curve where the heating effect was significant, and see evidence of this behavior toward the peaks of the three higher-field features. The effect of heating may also explain why the doublet structure of the 10~G $p$-wave resonances is not clear --- it has been washed out due to thermal broadening \cite{Ticknor2004}, the effect of which we can also see in the corresponding theoretical curve for a thermal sample at $1.9~\mu$K. While modifying the temperature used in our model gives values of $K_{21}$ that match more closely with the theory, there is still a discrepancy between experiment and theory in all but the $\simeq18$~G case.

\subsection{Relationship between dispersive signal and temperature}
\label{Sec:heatingDP}

The coupling factor $C_2$, entering the expression in \eref{Eq:DPs2} that relates the dispersive signal to the atomic population, depends on the geometry of both the atomic sample and the dispersive probe beam. In our experiments the dispersive probe beam parameters remain fixed, but the temperature increases slightly during Feshbach interactions, increasing the ensemble size according to \eref{Eq:sigT}. Because the temporal evolution of the temperature during these processes is unknown, we assumed a fixed coupling factor (measured at $T=1.4~\mu$K) to convert dispersive signal to atomic population.

The Rayleigh range of our dispersive beam is $z_{\mathrm{R}}\simeq 3$~mm $\gg \sigma_z \simeq 17~\mu$m, so we ignore intensity variations along the direction of propagation and consider only the Gaussian intensity profile in the radial ($xy$) plane,
\begin{equation}
I(x,y) = I_0 \mathrm{e}^{-\frac{2x^2+2y^2}{w_0^2}},
\end{equation}
where $I_0$ is the peak intensity and $w_0=28~\mu$m is the beam waist. Assuming our sample is at thermal equilibrium, the spatial profile of atoms can be well-approximated by a 2D Gaussian column density distribution
\begin{equation}
n(x,y) = 2 \sqrt{\pi}\sigma_zn_0\mathrm{e}^{-\left(\frac{x^2}{\sigma_x^2}+\frac{y^2}{\sigma_y^2 }\right)},
\end{equation}
where the factor of two arises from the fact that our sample consists of two identical Gaussian clouds in series along the direction of propagation, and $n_0$ is the peak atomic density
\begin{equation}
n_0 =\frac{N_2}{\sigma_x \sigma_y \sigma_z (2\pi)^{3/2}}.
\label{Eq:n0}
\end{equation}

\begin{figure}[!htbp]
	\centering
	\includegraphics[trim={0 0 0 0}, width=\columnwidth]{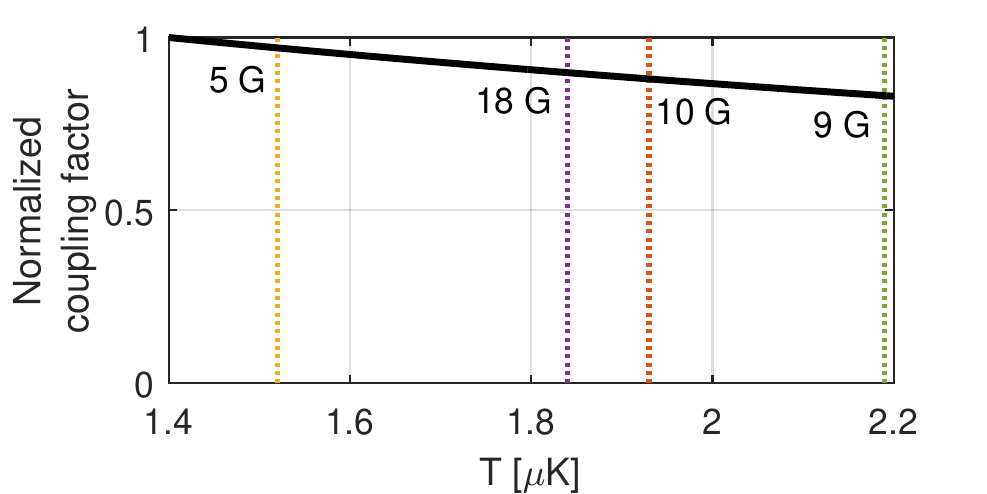}
	\caption{(Color online) Variation of geometric coupling factor with temperature, across the full temperature range of the samples in this work (solid line). The dotted lines indicate the maximum final temperatures measured for each of the four Feshbach resonances.}
	\label{fig:Coupling}
\end{figure}

The dispersive signal ($A$) can be expressed as
\begin{equation}
A =k \int_{-\infty}^{\infty} \int_{-\infty}^{\infty} n(x,y) I(x,y) dx dy,
\end{equation}
where $k$ is a fixed constant of proportionality that depends on the electronic system used to demodulate the signal, the resonant optical cross section, and the detuning of the probe light from resonance. Evaluating this integral gives
\begin{equation}
A = 2 k \pi^{3/2} I_0n_0 \frac{w_0^2\sigma_x\sigma_y\sigma_z}{(2\sigma_x^2+w_0^2)^{1/2}(2\sigma_y^2+w_0^2)^{1/2}}.
\end{equation}
To determine the dependence of the coupling factor on the temperature ($T$) and the atomic population ($N_2$), we substitute in \eref{Eq:sigT} and \eref{Eq:n0} to give
\label{eq:scat_xsec2}
\begin{align}
   A(T) =& \frac{k I_0}{\sqrt{2}}~\frac{w_0^2}{\left( w_0^2 + \frac{4k_BT}{m\omega_x^2} \right)^{1/2} \left( w_0^2 + \frac{4k_BT}{m\omega_y^2} \right)^{1/2}} ~N_2\nonumber\\
{=}&~ C_2(T) N_2.\label{Eq:ACN}
\end{align}

\Fref{fig:Coupling} shows the variation in the coupling factor over the full range of temperatures we observe, normalized to the value of the coupling factor at the initial temperature. The dotted lines indicate the maximum temperatures measured near each of the four Feshbach resonances, as recorded in \fref{fig:CompAIData}. Even for the maximum temperature increase (800~nK) encountered for the "on-resonance" value near 9~G, the expected correction is less than $20~\%$, and we stress that this is the worst case scenario; at all other magnetic fields and all earlier times the temperature increase is smaller, and thus the change in coupling factor is less pronounced. We conclude that the (approximate) conversion to atomic population $N_2$ from our dispersive signal $A$, based on a value for $C_2$ calibrated at a temperature of $T=1.4~\mu$K, is reasonable within our experimental setting.

%

\begin{thebibliography}{43}
\bibitem [{Sup()}]{noteTvsG}%
	\BibitemOpen
	\href@noop {} {}\bibinfo {note} {Units of gauss rather than tesla (the accepted SI unit for the magnetic field) have been used in this paper, as is conventional in this field of physics. The conversion is 1~G $=10^{-4}$~T.}\BibitemShut {Stop}%
\end{thebibliography}
\end{document}